\documentclass[aps,psfig,epsfig,preprint] {revtex4}
\usepackage{epsfig}
\usepackage{color}

\newcommand{\ppbar}{p\bar p}
\def \jp {J/\psi}

\newcommand {\sla}[1]{ #1 \!\!\!/}
\begin{document}
\draft
\title{
{\normalsize \hskip4.2in USTC-ICTS-05-15} \\{\bf Productions of
X(1835) as baryonium with sizable gluon content }}
\author{Gui-Jun Ding$^{a}$}
\author{Rong-Gang Ping$^{b}$}
\author{ Mu-Lin Yan$^{a}$}

\affiliation{\centerline{$^a$ Interdisciplinary Center for
Theoretical Study,} \centerline{University of Science and Technology
of China, Hefei, Anhui 230026, China} 
\centerline{$^b$Institute of High Energy Physics, P.O.Box 918(4),
Beijing 100049, China }}
\begin{abstract}
The $X(1835)$ has been treated as a baryonium with sizable gluon
content, and to be almost flavor singlet. This picture allows us to
rationally understand $X(1835)$ production in $J/\psi$ radiative
decays, and its large couplings with $p\overline{p}$,
$\eta^{\prime}\pi\pi$. The processes $\Upsilon(1S)\rightarrow \gamma
X(1835)$ and $J/\psi\rightarrow \omega X(1835)$ have been examined.
It has been found that $Br(\Upsilon(1S)\rightarrow\gamma
X(1835))Br(X(1835)\rightarrow p\overline{p})<6.45\times10^{-7}$,
which is compatible with  CLEO's recently experimental result
(Phys.Rev.$\mathbf{D73}$ (2006) 032001;hep-ex/0510015). The
branching fractions of $Br(J/\psi\rightarrow\omega X(1835))$,
$Br(J/\psi\rightarrow\rho X(1835))$ with $X(1835)\rightarrow
p\overline{p}$ and $X(1835)\rightarrow\eta^{\prime}\pi^{+}\pi^{-}$
have been estimated by the quark-pair creation model. We show that
they are heavily suppressed, so the signal of X(1835) is very
difficult, if not impossible, to be observed in these processes. The
experimental checks for these estimations are expected. The
existence of the baryonium nonet is conjectured, and a model
independent derivation of their production branching fractions is
presented.

\end{abstract}
\maketitle
\section{introduction}
Recently the BES collaboration has observed a new resonant state
X(1835) in the $\eta^{\prime}\pi\pi$ invariant mass spectrum in the
process $J/\psi\rightarrow\gamma\pi^{+}\pi^{-}\eta^{\prime}$
\cite{Bes1} with a statistical significance of 7.7$\sigma$. The fit
with the Breit-Wigner function yields   mass $\rm{M} = 1833.7\pm
6.1(stat)\pm 2.7(syst) \rm{MeV}/c^2$, width
$\Gamma=67.7\pm20.3(stat)\pm7.7(syst) \rm{MeV}/c^2$ and the product
branching fraction $Br(J/\psi\rightarrow\gamma
X(1835))Br(X(1835)\rightarrow\pi^{+}\pi^{-}\eta^{\prime})
=(2.2\pm0.4(stat)\pm0.4(syst))\times10^{-4}$. A narrow near
threshold enhancement in the proton-antiproton
(p$\overline{\rm{p}}$) mass spectrum was observed from the radiative
decay $J/\psi\rightarrow\gamma p\overline{p}$ \cite{Bes2}. This
enhancement can be fitted with either an S or P wave Breit-Wigner
resonance function. In the case of S-wave fit, the peak mass is
$\rm{M} = 1859^{+3}_{-10}(stat)^{+5}_{-25}(sys) \rm{MeV}/c^2$ with
the total width $\Gamma<30 \rm{MeV}/c^2$ at $90\%$ confidental level
and the product branching fraction $Br(J/\psi\rightarrow\gamma
X)Br(X\rightarrow p\overline{p})=(7.0\pm0.4^{+1.9}_{-0.8})\times
10^{-5}$.

The masses of the two structures observed in both $J/\psi\rightarrow
\gamma p\overline{p}$ and $J/\psi\rightarrow\gamma
\eta^{\prime}\pi^+\pi^-$ channels are overlap and $0^{-+}$ quantum
number for the resonance in $\eta'\pi^+\pi^-$ channel is possible. A
question arise if they are the same state, in Ref.\cite{Bes1} an
argument is presented if the final state interaction is included in
the fit of the $p\overline{p}$ mass spectrum, the width of the
resonance observed in $\gamma p\overline{p}$ channel will become
larger. Therefore, the X observed in both $p\overline{p}$ and
$\eta'\pi^+\pi^-$ channels could be the same state and it is named
as X(1835) in Ref.\cite{Bes1}. And this state couples strongly with
$\rm{p}\overline{p}$ and $\eta^{\prime}\pi^{+}\pi^{-}$, in the
recent talk of BES \cite{Bes4}, the estimation of
$Br(J/\psi\rightarrow\gamma X(1835))\sim(0.5-2)\times 10^{-3},
Br(X\rightarrow p\overline{p})\sim (4-14)\%$ are presented.
\par
However recently a negative experimental result has been reported by
CLEO collaboration\cite{Cleo}. They claimed that in the radiative
decay of $\Upsilon(1S)$ the narrow enhancement observed by BES near
$p\overline{p}$ mass threshold is not seen. The upper limit of the
product branching fraction for the decay $\Upsilon(1s)\rightarrow
\gamma X(1835)$, $X(1835)\rightarrow\gamma p\overline{p}$ has been
determined to be $Br(\Upsilon(1S)\rightarrow\gamma
X(1835))Br(X(1835)\rightarrow p\overline{p})<5\times
10^{-7}$\cite{Cleo}.

Moreover, another problem we would like to mention is that because
$Br(J/\psi\rightarrow\gamma X(1835))\sim(0.5-2)\times 10^{-3}$
claimed by BES in \cite{Bes1} is rather larger among $J/\psi$ decays
and $\omega$ is a photon like vector meson with negative  $G$ parity
, an experimental measurement of $Br(J/\psi\rightarrow\omega
X(1835))$ seems to be practicable in BES, or at least the signal of
$J/\psi\rightarrow\omega X(1835)$ should be seen in BES. However,
there are still not yet any results on this matter reported by BES,
therefore it is urgent to discuss the problem that whether the fact
that the signal of $J/\psi\rightarrow \omega X(1835)$ is not
revealed at the present stage contradicts  the existence of
$X(1835)$ or not.

In this case, the existence of $X(1835)$ seems to become a puzzle.
Therefore it is worth pursuing both the reasons why
$Br(\Upsilon(1S)\rightarrow\gamma X(1835))$ is so small that
$Br(\Upsilon(1S)\rightarrow\gamma X(1835))Br(X(1835)\rightarrow
p\overline{p})<5\times 10^{-7}$ and the reasons why there is still
not yet any information on $J/\psi\rightarrow\omega X(1835)$
reported by BES.  In this work we try to answer the above questions,
and try to illustrate that the absence of X(1835) signal from the
two  processes at present stage is due to the special structure of
X(1835).
\par
The theoretical interpretation of this exotic state is a great
challenge, and many proposals has been suggested
\cite{rosner,add,old,yuan,yan1,yan2,yan3,korea,hxg,lba,hzhu}. Some
of them interpret X(1835) as a $p\overline{p}$ bound state
\cite{add,yan1,yan2,yan3}, and large enough binding energy to bind
proton and antiproton together has been derived from the constitute
quark models\cite{yan3}. On the other hand, some authors identify
X(1835) as a pseudoscalar glueball \cite{korea,lba}, and in
Refs.\cite{yan2,hxg} the authors claim that there is large gluon
content in X(1835). Also some authors suggest that the two
structures observed in $J/\psi\rightarrow \gamma p\overline{p}$ and
$J/\psi\rightarrow\gamma \eta^{\prime}\pi^+\pi^-$ are not the same
state, and identify X(1835) as the $\eta$'s second radial excitation
\cite{hzhu}. Obviously, more theoretical and experimental efforts
are needed to determine whether X(1835) exists or not, and to be
sure that $X(1835)$ is a $p\overline{p}$ bound state or glueball or
something else. Motivated by solving the puzzles mentioned above and
getting the information about the structure of $X(1835)$ , we
investigate the productions of $X(1835)$ in $\Upsilon$ and $J/\psi$
decays in this work. The production of $X(1835)$ may provide
significant information on the structure of X(1835).

So far, the experiments strongly indicate that  X(1835) is almost
uniquely produced in $J/\psi$ radiative decays and it has large
coupling with $p\overline{p}$ and $\eta^{\prime}\pi\pi$. Whatever
X(1835) is a glueball or $p\overline{p}$ bound state or something
else, it must meet these two significant experimental facts. In this
work the possibility of X(1835) as a baryonium with sizable gluon
content are investigated. In this picture, the puzzles mentioned in
the above can be answered naturally.

The paper is organized as followings: In section II, we suggest
X(1835) as a baryonium with sizable gluon content, whose gluon
content is similar to that of$\eta^{\prime}$. In this picture, we
can easily understand the reasons why $\Upsilon(1S)\rightarrow\gamma
X(1835)$ and $J/\psi\rightarrow\omega X(1835)$ are not be seen at
the present stage. In section III we conjecture the existence of
pseudoscalar baryonium nonet and study its production in $J/\psi$
decay in a model independent way. Finally we briefly summary the
results and give some discussions.

\section{the possible structure of X(1835) and { $\Upsilon(1S)\rightarrow \gamma X(1835)$,
$J/\psi\rightarrow \omega X(1835)$}}

The production of X(1835) in $J/\psi$ radiative decay
$J/\psi\rightarrow\gamma\eta^{\prime}\pi^{+}\pi^{-}$ may indicate
that there is large gluon content in X(1835), as is shown in
Ref.\cite{yan2,hxg}. Also $J/\psi\rightarrow\gamma+gg,gg\rightarrow
hadrons$ provide an important search ground for the
glueball\cite{brodsky}, some people suggest that X(1835) is a
$0^{-+}$ glueball. However the lowest pseudoscalar glueball mass is
$2.1\sim 2.5 \rm{GeV}$ from the quenched lattice approach
\cite{lattice}, and $2.05\pm0.19 \rm{GeV}$ , $2.2\pm0.2 \rm{GeV} $
in QCD sum rules \cite{sumrule} and it seems difficult to explain
the large mass difference between 1835 MeV and the theoretical
prediction mass. On the other hand, even if X(1835) is a pure
glueball, it would mix with other mesonic states, such as
$\eta(1440)$, $\eta(1295)$ and $\eta_{c}(1S)$.
\par
Furthermore in Ref.\cite{yan3} we shown that X(1835) can be possibly
a baryonium and the relative large mass defect can be produced. In
Ref.\cite{yan2}, we pointed out that there is sizeable gluon content
in the Skyrmion-Baryonium $X(1860)$ (\textit{i.e.}, $X(1835)$) by
discussing the baryonium decay through baryon-antibaryon
annihilation in the Skyrme model. Distinguishing from the naive (or
old fashional) baryonium in the Fermi-Yang type
models\cite{old,FY,sakata,nakano}, the Skyrmion-Baryonium is
constructed in the model inspired by QCD, and therefore the gluon
inside the baryonium will play important role in the baryonium
physics, e.g, the baryonium decays and productions. Therefore, the
Skyrmion-Baryonium belongs to a sort of baryonium with sizable gluon
content. We address that in the naive baryonium model framework,  it
is difficult to simultaneously explain the large branching fraction
$X(1835)\rightarrow p\overline{p}$,
$X(1835)\rightarrow\eta^{\prime}\pi^{+}\pi^{-}$. The gluon content
in X(1835) should play essential role in the $X(1835)$
decay\cite{yan2}. So it is natural to treat X(1835) as a baryonium
with sizable gluon content, which looks like $\eta^{\prime}$ in some
sense, and mainly belongs to  a SU(3) flavor singlet.

In the following two subsections, we will start with this view to
examine the branching fractions of $\Upsilon(1S)\rightarrow\gamma
X(1835)$ and of $J/\psi\rightarrow\omega X(1835)$ respectively. We
will show that the branching fractions of both
$\Upsilon(1S)\rightarrow\gamma X(1835)$ and $J/\psi\rightarrow\omega
X(1835)$ are much smaller comparing to that of
$J/\psi\rightarrow\gamma X(1835)$. We will also predict the
branching fraction of $J/\psi\rightarrow\rho X(1835)$ is very small,
so we see that the process with visible $X(1835)$ may only be the
$J/\psi$ radiative decay at present stage.

\subsection{$\Upsilon(1S)\rightarrow\gamma X(1835)$}
According to Novikov et al.\cite{Novikov}, for the $J/\psi$
radiative decay, the photon is emitted by the $c$ quark with a
subsequent annihilation of the $c\overline{c}$ into light quarks
through the effect of the U(1)$_{A}$ anomaly . The creation of the
corresponding light quarks is controlled by the gluonic matrix
element
$\langle\frac{\alpha_{s}}{4\pi}G_{\mu\nu}\tilde{G}^{\mu\nu}|P_i\rangle$
($P_i$ is a pseudoscalar, it can be $\eta$ , $\eta^{\prime}$, and
X(1835) and so on ). Photon emission from the light quarks is
negligible as can be seen from the smallness of the
$J/\psi\rightarrow\gamma \pi$ decay width, this mechanism leads to
the following width for the $J/\psi$ radiative decay into the
pseudoscalar $P_i$
\begin{equation}
\label{1}\Gamma(J/\psi\rightarrow\gamma P_i)=\frac{2^5}{5^23^8}\pi
e_{c}^2\alpha^3_{em}K[J/\psi\gamma
P_i]^3(\frac{M_{J/\psi}}{m^2_{c}})^4\frac{|\langle\frac{\alpha_s}{4\pi}G_{\mu\nu}\tilde{G}^{\mu\nu}|P_i\rangle|^2}{\Gamma(J/\psi\rightarrow
e^{+}e^{-})}
\end{equation}
where $K[J/\psi\gamma P_i]$ is the momentum of the pseudoscalar
$P_i$ in the $J/\psi$ rest frame, and $K[J/\psi\gamma
P_i]=\frac{M_{J/\psi}}{2}(1-\frac{M^2_{P_i}}{M^2_{J/\psi}})$. Then
the ratio of the branching fractions between
$J/\psi\rightarrow\gamma X(1835)$ and $J/\psi\rightarrow\gamma
\eta^{\prime}$ is
\begin{equation}
\label{2}\frac{Br(J/\psi\rightarrow\gamma
X(1835))}{Br(J/\psi\rightarrow\gamma
\eta^{\prime})}=\frac{K[J/\psi\gamma X(1835)]^3}{K[J/\psi\gamma
\eta^{\prime}]^3}\frac{|\langle\frac{\alpha_s}{4\pi}G_{\mu\nu}\tilde{G}^{\mu\nu}|X(1835)\rangle|^2}
{|\langle\frac{\alpha_s}{4\pi}G_{\mu\nu}\tilde{G}^{\mu\nu}|\eta^{\prime}\rangle|^2}
\end{equation}
It is straightforward  to extend the anomaly dominance to the case
of the $\Upsilon(1S)$ radiative decay\cite{qwg,ma}. Then we have
\begin{equation}
\label{3}\frac{Br(\Upsilon(1S)\rightarrow\gamma
X(1835))}{Br(\Upsilon(1S)\rightarrow\gamma
\eta^{\prime})}=\frac{K[\Upsilon(1S)\gamma
X(1835)]^3}{K[\Upsilon(1S)\gamma
\eta^{\prime}]^3}\frac{|\langle\frac{\alpha_s}{4\pi}G_{\mu\nu}\tilde{G}^{\mu\nu}|X(1835)\rangle|^2}
{|\langle\frac{\alpha_s}{4\pi}G_{\mu\nu}\tilde{G}^{\mu\nu}|\eta^{\prime}\rangle|^2}.
\end{equation}
From Eq.(\ref{2}) and Eq.(\ref{3}), we get
\begin{eqnarray}
\nonumber Br(\Upsilon(1S)\rightarrow\gamma
X(1835))&=&\frac{K[\Upsilon(1S)\gamma
X(1835)]^3}{K[\Upsilon(1S)\gamma
\eta^{\prime}]^3}\frac{K[J/\psi\gamma
\eta^{\prime}]^3}{K[J/\psi\gamma
X(1835)]^3}\frac{Br(\Upsilon(1S)\rightarrow\gamma
\eta^{\prime})}{Br(J/\psi\rightarrow\gamma \eta^{\prime})}\\
\label{4}&&\times Br(J/\psi\rightarrow\gamma X(1835)).
\end{eqnarray}
For the $\Upsilon(1S)$ radiative decay
$\Upsilon(1S)\rightarrow\gamma\eta^{\prime}$, only upper limit has
been obtained, which is
$Br(\Upsilon(1S)\rightarrow\gamma\eta^{\prime})<1.6\times 10^{-5}$
at 90\% confidental level\cite{pdg}. Using again the Particle Data
Group's value
$Br(J/\psi\rightarrow\gamma\eta^{\prime})=(4.31\pm0.30)\times10^{-3}$\cite{pdg},
and substituting it into Eq.(\ref{4}), we obtain
\begin{equation}
\label{5}Br(\Upsilon(1S)\rightarrow\gamma
X(1835))<9.22\times10^{-3}Br(J/\psi\rightarrow\gamma X(1835)).
\end{equation}
Thus, since BES has already determined $Br(J/\psi\rightarrow\gamma
X(1835))Br(X(1835)\rightarrow
p\overline{p})=(7.0\pm0.4^{+1.9}_{-0.8})\times10^{-5}$\cite{Bes2},
and by eq.(\ref{5}), we finally get a reasonable  estimation:
\begin{equation}
\label{6}Br(\Upsilon(1S)\rightarrow\gamma
X(1835))Br(X(1835)\rightarrow p\overline{p})<6.45\times10^{-7}.
\end{equation}
This estimation is compatible with CLEO collaboration's result of
$Br(\Upsilon(1S)\rightarrow\gamma X(1835))Br(X(1835)\rightarrow
p\overline{p})<5\times10^{-7}$. So it is not surprising that the
CLEO collaboration don't see the signal of X(1835) in the
$\Upsilon(1S)$ radiative decay, and it doesn't mean that X(1835)
seen by BES is an experimental artifact.

\subsection{$J/\psi\rightarrow\omega X(1835)$}

In this subsection, we examine the branching fraction of
$J/\psi\rightarrow\omega X(1835)$. Since $Br(J/\psi\rightarrow
\gamma X(1835))$ is rather larger\cite{Bes1,Bes2}, one could expect
$Br(J/\psi\rightarrow \omega X(1835))$ may also be reasonably large
too, or at least be visible at present stage. In this way the
existence of $X(1835)$ may be rechecked in the non-radiative decay
channel of $J/\psi$. However, this is  only a naive conjecture, and
there is not yet any data on this branching fraction experimentally.
So a theoretical estimation on $Br(J/\psi\rightarrow \omega
X(1835))$ is necessary. Our estimations in this subsection are still
based on the baryonium picture discussed in the above.


Unlike the radiative decays $J/\psi\rightarrow\gamma X(1835)$ and
$\Upsilon(1S)\rightarrow\gamma X(1835)$, where the gluon component
plays important role due to the $U_A(1)$ anomaly. For the decay
$J/\psi\rightarrow\omega X(1835)$, the processes to which the
$U_A(1)$ anomaly contributes are suppressed, and the baryonic
component dominates this process, the same is true for
$J/\psi\rightarrow\omega\eta'$ . We think the decay process
$J/\psi\rightarrow \omega X(1835)$ proceeds via two steps as
illustrated in Fig.1. In first step, the $c\overline{c}$ pair
annihilate into three gluons, followed by the materialization of
each gluon into a pair of quark-antiquark, this process can be
calculated from perturbative QCD to the lowest order. Also a pair of
quark-antiquark are created from the vacuum, and this process can be
described by the quark pair creation model (the $^3P_0$ model). Then
in the second step the quarks and the antiquarks combine to form
$\omega$ and X(1835). Here, the nonperturbative dynamics is included
by the hadron's wave functions in the naive quark model.

The quark pair creation model which describes the process that a
pair of quark-antiquark with quantum number $J^{PC}=0^{++}$ is
created from vacuum was first proposed by Micu\cite{Micu} in 1969.
In the 1970s, this model was developed by Yaouanc et al.
\cite{qpc,yaouanc,yaouanc-1,yaouanc-book} and applied to study
hadron decays extensively. The $^3P_0$ quark pair creation model has
proven to be a successful mechanism for describing strong decay of
light mesons \cite{isgur,qpc-90,ackleh}. It also has been shown that
the $^3P_0$ quark pair creation mechanism may play important role
for some exclusive decay in the charmonium
sector\cite{ping1,ping2,ping3}. In the $^3P_0$ model, the created
quark pairs with any color and any flavor can be generated anywhere
in space, but only those whose color-flavor wave functions and
spatial wave functions overlap with those of outgoing hadrons can
make a contribution to the final decay width. The hamiltonian for
creating a quark pair can be defined in the $^3P_0$ model in terms
of quark and antiquark creation operators $b^+$ and $d^{+}$
\cite{ackleh},
\begin{equation}
H_{I}=\sum_{i,j,\alpha,\beta,s,s^{\prime}}\int d^3k~
g_I[\overline{u}(\mathbf{k},s)v(-\mathbf{k},s^{\prime})]b^{+}_{\alpha,i}
(\mathbf{k},s)d^{+}_{\beta,j}(-\mathbf{k},s^{\prime})\delta_{\alpha\beta}\hat{C}_{I}
\end{equation}
where $\alpha(\beta)$ and i(j) are the flavor and color indices of
the created quarks (antiquarks), and $u(\mathbf{k}, s)$ and
$v(\mathbf{k}^{\prime}, s^{\prime})$ are free Dirac spinors for
quarks and antiquarks respectively. $\hat{C}_I=\delta_{ij}$ is the
color operator for $q\overline{q}$ and $g_{I}$ is the strength of
the decay interaction, which is assumed as a constant in these
processes. In the nonrelativistic limit, $g_I$ can be related to
$\gamma$, the strength of the conventional $^3P_0$ model, by $g_I =
2m_q\gamma$\cite{ackleh}. In order to cancel the large uncertainty
in $g_I$ and the overall constant dependence, we will calculate the
ratio $\frac{\Gamma(J/\psi\rightarrow\omega
X(1835))}{\Gamma(J/\psi\rightarrow\omega\eta^{\prime})}$ in the
following. The process $J/\psi\rightarrow\omega\eta^{\prime}$ is
schematically shown in Fig.2, where the electromagnetic decay
process is not shown. For the $J/\psi$ decaying into hadrons, the
ratio between the hadronic decay width and the electromagnetic decay
width is about 5 \cite{decay}, i,e, $\frac{\Gamma(J/\psi\rightarrow
ggg\rightarrow hadrons)}{\Gamma(J/\psi\rightarrow \gamma\rightarrow
hadrons)}\simeq5$. Thus we can include the contribution of the
electromagnetic decay to $J/\psi\rightarrow\omega\eta^{\prime}$
though the above ratio.

For the $\jp\to PV$ decays,  the parity transformation is conserved,
and the transitional amplitude square is
$\sum_\Lambda|M(\Lambda)|^2=(1+cos^2\theta) |A_1|^2$, where
$\Lambda$ is the $\jp$ helicity, which is taken as $\Lambda= \pm 1$
if it is produced from $e^+e^-$ annihilations, $A_1$ is the helicity
amplitude with vector meson helicity equaling to 1, and $\theta$ is
the polar angle of the outgoing meson. The decay width
$\Gamma={|\mathbf{P}|\over 6\pi M_{\jp}^2}|A_1|^2 $, here
$\mathbf{P}$ is the momentum of out-going mesons.
\subsubsection{$\jp\to\omega X(1835)\to \omega\ppbar $}
The color factors for the fig.1 are:
\begin{itemize}
\item color factor for fig1.(a), $c_a={5\over 54}$
\item color factor for fig1.(b), $c_b={5\over 432}$ (to be negligible)
\end{itemize}
The amplitude can be obtained according to the standard Feynman
rules with the quark pair creation hamiltonian included, which is
expressed as followings:
\begin{eqnarray}
T_{\Lambda,s}(J/\psi\rightarrow\omega X(1835))&&=
C_0c_a\alpha_s^{3/2} <\phi_\omega\phi_{X}|\overline
u(\mathbf{p}_1,s_1)\gamma_\mu v(\mathbf{q}_1,\bar s_1)\overline{u}(\mathbf{p}_2,s_2)\gamma_\nu v(\mathbf{q}_2,\bar s_2)\overline{u}(\mathbf{p}_3,s_3)\gamma_\rho \nonumber \\
&&v(\mathbf{q}_3,\bar
s_3)\epsilon_{\psi}^{(\Lambda)\lambda}\times{g_{\mu\lambda}g_{\nu\rho}+g_{\nu\lambda}g_{\mu\rho}+g_{\rho\lambda}g_{\mu\nu}\over
k_1^2k_2^2k_3^2}g_I\overline{u}(\mathbf{p}_4,s_4)v(\mathbf{q}_4,\bar
s_4)|\phi_{\jp}>
\end{eqnarray}
where $C_0$ is coupling constant for $\jp\to ggg$, and $\alpha_s$ is
the the strong coupling constant. $k_i$ is the gluonic momentum, and
the normalization of Dirac spinor is taken as $\overline{u}
u=-\overline{v} v=m/E$. $\phi_{\omega},\phi_X$ and $\phi_{\jp}$
represent the wave functions of $\omega, X(1835)$ and $\jp$
respectively. And the helicity amplitude is:
\begin{eqnarray}\label{}
A_{\Lambda,s}(J/\psi\rightarrow\omega X(1835))&&=\int
\prod_{i=1,4}{d^3\mathbf{q}_i\over (2\pi)^3}{d^3\mathbf{p}_i\over
(2\pi)^3} {d^3\mathbf{t}_1\over (2\pi)^3}{d^3\mathbf{t}_2\over
(2\pi)^3}~~ T_{\Lambda,s}(J/\psi\rightarrow\omega
X(1835))\nonumber\\
&&\times(2\pi)^3\delta^3(\mathbf{p}_\omega-\mathbf{p}_1-\mathbf{q}_4)(2\pi)^3\delta^3
(\mathbf{p}_2+\mathbf{p}_3
+\mathbf{p}_4-\mathbf{t}_1)\nonumber\\
&&(2\pi)^3\delta^3(\mathbf{q}_1+\mathbf{q}_2
+\mathbf{q}_3-\mathbf{t}_2)\times(2\pi)^3\delta^3(\mathbf{t}_1+\mathbf{t}_2-\mathbf{p}_X)\nonumber\\
&&(2\pi)^3\delta^3(\mathbf{p}_4+\mathbf{q}_4)
\end{eqnarray}
Here $\Lambda$ and $s$ denote the helicity of $\jp$ and $\omega$
respectively.
\subsubsection{$\jp\to \omega\eta'$}
The color factors for the fig.2 are:
\begin{itemize}
\item color factor for fig2 (a): $c'_a={5\sqrt 3\over 54}$.
\item color factor for fig2 (b): $c'_b={5\sqrt 3\over 216}$ (negligible).
\end{itemize}
The corresponding helicity amplitude is
\begin{eqnarray}
A_{\Lambda,s}(J/\psi\rightarrow\omega\eta')&=&\int{d^4 k_1\over
(2\pi)^4}{d^4 k_2 \over (2\pi)^4}{d^3\mathbf{p}_1\over
(2\pi)^3}{d^3\mathbf{p}_2\over (2\pi)^3}{d^3\mathbf{q}_1\over
(2\pi)^3}T_{\Lambda,s}(J/\psi\rightarrow\omega\eta')\nonumber\\
&&\times(2\pi)^3\delta^3(\mathbf{q}_\omega-\mathbf{p}_1+\mathbf{p}_2)(2\pi)^3\delta^3(\mathbf{q}_{\eta'}-\mathbf{q}_1-\mathbf{p}_2)
\end{eqnarray}
where $T_{\Lambda,s}(J/\psi\rightarrow\omega\eta')$ is the
followings,
\begin{eqnarray}
T_{\Lambda,s}(J/\psi\rightarrow\omega\eta')&=&C_0c'_a\alpha_s^{3/2}<\phi_\omega(q_\omega,s)\phi_{\eta'}(q_{\eta'})|\overline{u}
(p_1,s_1)\gamma_\mu{1\over \sla{p}_1-\sla{k}_1-m}\gamma_\nu{1\over
\sla{q}_1-\sla{k}_2-m}\gamma_\rho v(q_1,\bar s_1) \nonumber\\
&\times&\epsilon_\psi^{(\Lambda)\lambda}{g_{\mu\lambda}g_{\nu\rho}+g_{\nu\lambda}g_{\mu\rho}+g_{\rho\lambda}g_{\mu\nu}\over
k_1^2k_2^2k_3^2}\nonumber g_I\overline{u}(p_2,s_2)v(-p_2,\bar s_2
)|\phi_{\jp}>\\
\end{eqnarray}
For simplicity, we make use of the on-shell approximation,
\textit{i.e.,} ${1\over k_1^2k_2^2}\to
-2\pi^2\delta(k_1^2)\delta(k_2^2)$, and with the replacement:
\begin{equation}
\int{d^4 k_1d^4 k_2 \over k_1^2k_2^2k_3^2}\to-{\pi^2\over
2}\int_0^{M_\psi}dk_1^0\int_0^{M_\psi-k_1^0}dk_2^0\int
d\Omega_1d\Omega_2{k_1^0k_2^0 \over
M_\psi^2-2k_1^0M_\psi-2k_2^0M_\psi+2k_1\cdot k_2}
\end{equation}
\subsubsection{Numerical results}
The spin-flavor wave functions of the mesons $\omega$ and
$\eta^{\prime}$ are well-known in quark model, and the spatial wave
function is taken as the simple harmonic oscillator wave function,
\textit{i.e.,} $\phi(\mathbf{k})={(2\pi)^{3/2}\over
(\pi\beta^2)^{3/4}}e^{-\mathbf{k}^2/2\beta^2}$. Since X(1835) is
assumed as a baryonium with $J^{PC}=0^{-+}$, it's spatial wave
function is the product of the proton spatial wave function,
antiproton spatial wave function and the relative spatial wave
function between them. It's expressed as followings,
\begin{equation}\phi_X=\phi_p(\mathbf{p}_\rho,\mathbf{p}_\lambda)\phi_{\overline{p}}(\mathbf{q}_\rho,\mathbf{q}_\lambda)\phi_{p\overline{p}}(\mathbf{t}_1-\mathbf{t}_2)
\end{equation}
where $\phi_p(\mathbf{p}_\rho,\mathbf{p}_\lambda)={(2\pi)^{3/2}\over
(\pi\beta^2)^{3/2}}e^{-(\mathbf{p}_\rho^2+\mathbf{p}_\lambda^2)/2\beta^2}$,
and $\phi_{p\overline{p}}(\mathbf{t}_1-\mathbf{t}_2)$ is of the same
formalism as the simple harmonic oscillator wave function. $
\textrm{with}\;\;\mathbf{p}_\rho={1\over \sqrt
6}(\mathbf{p}_2+\mathbf{p}_3-2\mathbf{p}_4),\;\;
\mathbf{q}_\rho={1\over \sqrt
6}(\mathbf{q}_1+\mathbf{q}_2-2\mathbf{q}_3),\;\;
\mathbf{p}_\lambda={1\over \sqrt 2}(\mathbf{p}_2-\mathbf{p}_3),\;\;
\mathbf{q}_\lambda={1\over \sqrt
2}(\mathbf{q}_1-\mathbf{q}_2),\;\;\mathbf{t}_1=\mathbf{p}_2+\mathbf{p}_3+\mathbf{p}_4\;\;\rm{and}\;\;
\mathbf{t}_2=\mathbf{q}_1+\mathbf{q}_2+\mathbf{q}_3$. The
spin-flavor wave function is the followings,
\begin{eqnarray}
\nonumber && {1\over 2\sqrt 2}\{
[\chi_p^\rho(\uparrow)\phi_p^\rho+\chi_p^\lambda(\uparrow)\phi_p^\lambda]
[\chi_{\overline{p}}^\rho(\downarrow)\phi_{\overline{p}}^\rho+\chi_{\overline{p}
}^\lambda(\downarrow)\phi_{\overline{p}}^\lambda]\\
&&-[\chi_p^\rho(\downarrow)\phi_p^\rho+\chi_p^\lambda(\downarrow)\phi_p^\lambda][\chi_{\overline{p}
}^\rho(\uparrow)\phi_{\overline{p}}^\rho+\chi_{\overline{p}}^\lambda(\uparrow)\phi_{\overline{p}}^\lambda]\}
\end{eqnarray}
where $\phi_{p}^{\rho}$ and $\phi_{p}^{\lambda}$ are $\rho$-type and
$\lambda$-type nucleon flavor wave function respectively, and
similarly for $\chi_{p/\overline{p}}^{\rho}$ and
$\chi_{p/\overline{p}}^{\lambda}$ .

There are four parameters to be determined in our
calculation,\textit{i.e.,} the quark mass $m_u,m_d$, the harmonic
oscillation parameter $\beta$ for hadrons and X(1835). The quark
mass are taken as $m_u=m_d=0.31\rm{GeV}$. In most calculations in
quark model, the harmonic oscillation parameter is fitted to the
hadron decay width, which gaves $\beta=0.4$GeV \cite{isgur,ackleh}.
The harmonic oscillation parameter of X(1835) is determined by
assuming that the radius of $p\overline{p}$ is about $1\sim 2$fm,
which corresponds to the the parameter $\beta_X=0.15\sim 0.30$GeV.
The ratio of $\Gamma(J/\psi\rightarrow\omega
X(1835))/\Gamma(J/\psi\rightarrow\omega\eta')$ is calculated in
terms of different set of harmonic oscillation parameters $\beta$
and $\beta_X$ as listed in Table 1. It is clear to see that the
ratio is very sensitive the parameter $\beta$ and not sensitive to
$\beta_X$.

\begin{table}[htbp]
 \centering
 \caption{Numerical results of $\frac{\Gamma(J/\psi\rightarrow\omega X(1835))}{\Gamma(J/\psi\rightarrow\omega\eta')}$ corresponding to the different set of harmonic oscillation parameters $\beta$ and $\beta_X$, with the contribution of the electromagnetic process included , where the quark mass are taken as $m_u=m_d=0.31$GeV.}\label{}
\begin{tabular}{|c|c|c|c|c|c|}
  \hline\hline
  $\beta$(GeV)& \multicolumn{4}{c}{$\beta_X$(GeV)}\vline&$\rm{Average\;\; of}\;\; \frac{\Gamma(J/\psi\rightarrow\omega X)}{\Gamma(J/\psi\rightarrow\omega\eta')}$ \\ \cline{2-5}
    & 0.15 & 0.20 & 0.25 &0.30&  \\\hline
0.36&6.2&5.9&5.0&4.1&$5.2\pm 1.0$\\
0.40&0.15&0.13&0.12&0.10&$0.12\pm 0.02$\\
0.46&0.006&0.006&0.005&0.005&$0.004\pm0.0008$\\
0.52&$5.9\times 10^{-4}$&$5.6\times 10^{-4}$&$4.8\times 10^{-4}$&$4.0\times 10^{-4}$&$(5.0\pm 0.9)\times 10^{-4}$\\
  \hline
\end{tabular}
\end{table}

As most quark model studies on the meson decays, we use the
parameters $m_u=m_d=0.31,\beta=0.4$GeV as our favorable parameters.
In our calculation, the uncertainties are from the parameter
$\beta_X$, the ignored decay modes depressed by color factor and the
accuracy of the numerical calculation. From our estimation, the
uncertainty of the $\beta_X$ within our setting range is about 20\%.
The contribution from the fig.1(b) and fig.2(b) are of the same
order as that from fig.1(a) and fig.2(a) respectively. The color
depressed decay modes will bring in uncertainty of about 6\%, the
uncertainty of the numerical evaluation is about 8\%, then the total
uncertainty is about 22\%. Including these uncertainties, we predict
the ratio
$\Gamma(J/\psi\rightarrow\omega\eta')/\Gamma(J/\psi\rightarrow\omega
X)=0.12\pm 0.02$. Using the PDG value $Br(\jp\to
\omega\eta')=(1.67\pm 0.25)\times 10^{-4}$, we predict that
$Br(\jp\to \omega X(1835))=(2.00\pm 0.35)\times 10^{-5}$.

Comparing this result with $Br(J/\psi\rightarrow\gamma X(1835))\sim
(0.5-2)\times 10^{-3}$\cite{Bes1}, we see that the production rate
of $X(1835)$ in the process $J/\psi\rightarrow\omega X(1835)$ is
less than that in $J/\psi\rightarrow\gamma X(1835)$ about two
orders. Therefore, it is not surprising that the signal of
$J/\psi\rightarrow\omega X(1835)$ has not be seen by BES or other
laboratories so far. In other words, the absence of the signal in
the decay $J/\psi\rightarrow\omega X(1835)$ at present can not be
thought as an evidence against the existence of $X(1835)$ .

Using BES's estimations of $Br(X(1835)\rightarrow
p\overline{p})\sim(4-14)\%$ and
$Br(X(1835)\rightarrow\eta^{\prime}\pi^{+}\pi^{-})\sim 3
Br(X(1835)\rightarrow p\overline{p})$\cite{Bes1,Bes4}, we get
further two useful estimations about the product branching
fractions:
\begin{eqnarray}
8.00\times10^{-7}<Br(J/\psi\rightarrow\omega
X(1835))Br(X(1835)\rightarrow p\overline{p})<2.80\times10^{-6},\\
\label{14}2.40\times10^{-6}<Br(J/\psi\rightarrow\omega
X(1835))Br(X(1835)\rightarrow
\eta^{\prime}\pi^{+}\pi^{-})<8.40\times10^{-6}.
\end{eqnarray}
Comparing them with the data\cite{Bes1} $Br(J/\psi\rightarrow\gamma
X)Br(X\rightarrow p\overline{p})=(7.0\pm0.4^{+1.9}_{-0.8})\times
10^{-5}$ and $Br(J/\psi\rightarrow\gamma
X(1835))Br(X(1835)\rightarrow\pi^{+}\pi^{-}\eta^{\prime})
=(2.2\pm0.4(stat)\pm0.4(syst))\times10^{-4}$ respectively, we see
also both $ Br(J/\psi\rightarrow\omega X(1835))Br(X(1835)\rightarrow
p\overline{p})$ and $ Br(J/\psi\rightarrow\omega
X(1835))Br(X(1835)\rightarrow \eta^{\prime}\pi^{+}\pi^{-})$ are also
very small. So the signal of X(1835) is very difficult, if not
impossible, to be observed in the process $J/\psi\rightarrow\omega
X(1835)$ with $X(1835)\rightarrow p\overline{p}$ or
$X(1835)\rightarrow \eta'\pi^{+}\pi^{-}$.

Finally we discuss the production of X(1835) in the process
$J/\psi\rightarrow\rho X(1835)$. In this process the G-parity is not
conserved, and it proceeds  through a virtual photon
$c\overline{c}\rightarrow\gamma^{*}$. Contributions from the
isospin-violating part of QCD are supposedly very small. Furthermore
the masses of $\rho$ and $\omega$ are approximately equal, so
$Br(J/\psi\rightarrow\rho X(1835))<Br(J/\psi\rightarrow\omega
X(1835))$ (the same holds true for $J/\psi\rightarrow\omega
\eta^{\prime}$ and $J/\psi\rightarrow\rho \eta^{\prime}$, that is
$Br(J/\psi\rightarrow\omega \eta^{\prime})>Br(J/\psi\rightarrow\rho
\eta^{\prime})$ ). This means that we also can not see X(1835) in
the process $J/\psi\rightarrow\rho X(1835)$.

After the above analysis, we conclude that comparing to $Br(J/\psi
\rightarrow \gamma X(1835))$, the branching fractions of $J/\psi
\rightarrow V X(1835)$ with $V~\rm{being}~\omega~\rm{or }~\rho $ are
heavily suppressed due to its special structure. The search for the
X(1835) in these decays seems impossible at the present stage.

\section{ baryonium nonet and its production in a model independent way}
The BES collaboration has observed not only the $p\overline{p}$
enhancement\cite{Bes1,Bes2}, but also the $p\overline{\Lambda}$
enhancement\cite{Bes5}. These two states can belong to flavor
$\mathbf{1}-plet$,~$\mathbf{8}-plet$,~$\mathbf{10}-plet$,~$\mathbf{\overline{10}}-plet$,~or
$\mathbf{27}-plet$. It seems that at least a baryonium nonet exists.
Theoretically, we have predicted existence of such  a baryonium
nonet  in Ref.\cite{yan3}. The baryonium nonet was also suggested
from the Fermi-Yang-Sakata model in Ref.\cite{yuan}. The nonet can
be pseudoscalar  or  vector multiplet\cite{yuan,yan3}, and the
corresponding weight diagram is shown in Fig.3 and Fig.4. The
pseudoscalar and vector enhancement octet are respectively denoted
by $E_{P_i}$ and $E_{V_i} (i=1\cdot\cdot\cdot 8)$ as follows
\begin{eqnarray}
\nonumber&&E_{\pi^{\pm}}=\frac{1}{\sqrt{2}}(E_{P_1}\mp iE_{P_2}),~
E_{\pi^{0}}=E_{P_3},~E_{K^{\pm}}=\frac{1}{\sqrt{2}}(E_{P_4}\mp
iE_{P_5})\\
\label{19}&&E_{K^{0}}=\frac{1}{\sqrt{2}}(E_{P_6}-iE_{P_7}),~E_{\overline{K}^0}=\frac{1}{\sqrt{2}}(E_{P_6}+iE_{P_7}),~E_{\eta_8}=E_{P_8}
\end{eqnarray}
It is useful to add the singlet to the octet $E_{P}$ by defining
$E_{\eta_1}=E_{P_0}$, thereby creating the nonet
$E_P=(E_{P_0},E_{P_i})$. If the pseudoscalar glueball and radially
exciting states are ignored, the physics states $E_{\eta^{\prime}}$
and $E_{\eta}$ are mixing of $E_{\eta_8}$ and $E_{\eta_0}$ with the
mixing angle $\varphi_P$
\begin{equation}
\label{20}E_{\eta_8}=\cos\varphi_P E_{\eta}+\sin\varphi_P
E_{\eta^{\prime}},~~E_{\eta_1}=-\sin\varphi_P E_{\eta}+\cos\varphi_P
E_{\eta^{\prime}}
\end{equation}
Similarly for the vector enhancement nonet
$E_{V}=(E_{\omega_1},E_{V_i})$, the physics states $E_{\omega}$ and
$E_{\phi}$ are mixing of $E_{\omega_8}$ and $E_{\omega_1}$ with the
mixing angle $\varphi_V$
\begin{equation}
\label{21}E_{\omega_8}=\cos\varphi_V E_{\phi}+\sin\varphi_V
E_{\omega},~~E_{\omega_1}=-\sin\varphi_V E_{\phi}+\cos\varphi_P
E_{\omega}
\end{equation}
We identify the $p\overline{p}$ enhancement X(1835) as the states
$E_{\eta^{\prime}}$, while the $p\overline{\Lambda}$ enhancement
should be $E_{K^+}$ or $E_{K^{*+}}$, and $E_{K^{*+}}$ is favored
over $E_{K^+}$ from the analysis of Ref.\cite{yan3}.
\par
We can consider flavor SU(3) breaking by choosing a nonet pointing
to a fixed direction in SU(3) space particularly for the desired
breaking. We will consider two types of SU(3) breaking, first SU(3)
is broken due to $m_s\neq m_u, m_d$($m_u=m_d$ is assumed), the quark
mass term is
$m_d(d\overline{d}+u\overline{u})+m_ss\overline{s}=m_0q\overline{q}+\sqrt{\frac{1}{3}}(m_d-m_s)\overline{q}\lambda_8q$,
where $q=(u,d,s)$ and $m_0=\frac{1}{3}(2m_d+m_s)$. We can see this
SU(3) breaking corresponds to a nonet $\mathbf{M}$, pointing to the
8th direction, \textit{i.e.,} $M^a=\delta^{a8}$. Second, the
electromagnetic effects violate SU(3) invariance, since the photon
coupling to quarks is
$\frac{2}{3}\overline{u}\gamma_{\mu}u-\frac{1}{3}\overline{d}\gamma_{\mu}d-\frac{1}{3}\overline{s}\gamma_{\mu}s=
\frac{1}{2}\overline{q}\gamma_{\mu}(\lambda_3+\frac{\lambda_8}{\sqrt{3}})q$.
This symmetry breaking effect corresponds to a nonet $\mathbf{E}$,
given by $E^a=\delta^{a3}+\sqrt{\frac{1}{3}}\delta^{a8}$. In the
following we consider the process of $J/\psi\rightarrow E_{P} P$,
that means $J/\psi$ decays into a pseudoscalar baryonium
($E_{\pi},E_{K},E_{\eta},E_{\eta^{\prime}}$) and a pseudoscalar
($\pi,K,\eta,\eta^{\prime}$), the process of $J/\psi\rightarrow
E_{P} V$, \textit{i.e.,} $J/\psi$ decays to a pseudoscalar baryonium
($E_{\pi},E_{K},E_{\eta},E_{\eta^{\prime}}$) and a vector
($\rho,K^*,\omega,\phi$) and the process of $J/\psi\rightarrow E_{V}
P$, \textit{i.e.,} $J/\psi$ decays to a vector baryonium
($E_{\rho},E_{K^*},E_{\omega},E_{\phi}$) and a pseudoscalar
($\pi,K,\eta,\eta^{\prime}$) in a model independent way via SU(3)
symmetry with the effects of electromagnetic and mass breaking of
the SU(3) symmetry being considered\cite{haber1,haber2}. Since the
phase space factor is proportional to the cube of the final
three-momentum, we define the reduced branching fraction
$\widetilde{Br}(J/\psi\rightarrow E_P P)=Br(J/\psi\rightarrow E_P
P)/P^3_P$, here $P_P$ is the momentum of the pseudoscalar P in the
$J/\psi$ rest frame, and the reduced branching fractions
$\widetilde{Br}(J/\psi\rightarrow E_P V)$,
$\widetilde{Br}(J/\psi\rightarrow E_V P)$ are defined in the same
way.
\subsection{$J/\psi\rightarrow E_{P}P $}
These processes occur completely due to the SU(3) breaking effects,
we can constructed charge conjugation invariant and SU(3) invariant
effective lagrangian involving the symmetry breaking nonet
$\mathbf{E}$ or $\mathbf{M}$, which may be written as followings,
\begin{equation}
\label{22}{\cal{L}}_{eff}=f^{abc}\Psi^{\mu}E_{P_a}\stackrel{\leftrightarrow}{\partial_{\mu}}P_b(g_MM^c+g_EE^c)
\end{equation}
Here the new parameter $g_M$ and $g_E$ parametrize the SU(3)
breaking effects. From Eq.(19) and Eq.(22) we can obtain the
following reduced branching fractions
\begin{eqnarray}
\nonumber&&\widetilde{Br}(J/\psi\rightarrow
E_{\pi^+}\pi^-)=\widetilde{Br}(J/\psi\rightarrow
E_{\pi^-}\pi^+)=|g_E|^2\\
\nonumber&&\widetilde{Br}(J/\psi\rightarrow
E_{K^+}K^-)=\widetilde{Br}(J/\psi\rightarrow
E_{K^-}K^+)=|\frac{\sqrt{3}}{2}g_M+g_E|^2\\
\nonumber&&\widetilde{Br}(J/\psi\rightarrow
E_{K^0}\overline{K}^0)=\widetilde{Br}(J/\psi\rightarrow
E_{\overline{K}^0}K^0)=\frac{3}{4}|g_M|^2\\
\label{23}&&\widetilde{Br}(J/\psi\rightarrow
E_{\pi^0}\pi^0)=\widetilde{Br}(J/\psi\rightarrow
E_{\eta}\pi^0)=\widetilde{Br}(J/\psi\rightarrow
E_{\eta^{\prime}}\pi^0)=0
\end{eqnarray}
The last formula means that the process $J/\psi\rightarrow\pi^0
X(1835)$, $X(1835)\rightarrow p\overline{p}$ should be forbidden,
which is indeed not observed \cite{Bes2}, and it is forbidden
because of C parity.
\subsection{$J/\psi\rightarrow E_P V$}
Following the same way as in the discussion of $J/\psi\rightarrow
E_P P$, we construct the charge conjugation invariant, SU(3)
invariant effective lagrangian including the symmetry breaking nonet
$\mathbf{E}$ and $\mathbf{M}$ , which may be written as follows:
\begin{eqnarray}
&&\nonumber{\cal
L}_{eff}=\varepsilon_{\mu\nu\alpha\beta}F^{\mu\nu}_{\Psi}\{g_8F^{\alpha\beta}_{V_a}E_{P_a}+g_1F^{\alpha\beta}_{\omega_1}E_{\eta_1}
+[g_{M,88}d^{abc}F^{\alpha\beta}_{V_a}E_{P_b}M^c+\sqrt{\frac{2}{3}}g_{M,18}F^{\alpha\beta}_{V_a}M^aE_{\eta_1}\\
\nonumber&&+\sqrt{\frac{2}{3}}g_{M,81}F^{\alpha\beta}_{\omega_1}M^aE_{P_a}]+[g_{E,88}d^{abc}F^{\alpha\beta}_{V_a}E_{P_b}E^c+\sqrt{\frac{2}{3}}g_{E,18}F^{\alpha\beta}_{V_a}E^aE_{\eta_1}
+\sqrt{\frac{2}{3}}g_{E,81}F^{\alpha\beta}_{\omega_1}E^aE_{P_a}]\}\\
\end{eqnarray}
Where $F^{\mu\nu}_{\Psi}$ is the strength of the $J/\psi$ field with
$F^{\mu\nu}_{\Psi}=\partial^{\mu}\Psi^{\nu}-\partial^{\nu}\Psi^{\mu}$,
and $F^{\alpha\beta}_{\omega_1}$, $F^{\alpha\beta}_{V_a}$ are
respectively the field strength of the vector field $\omega_1$ and
$V_a$. We assume nonet symmetry holds true within a reasonable
approximation, which relate the octet to the singlet, then the
relations $g_8=g_1\equiv g$, then we have the relations
$g_{M,88}=g_{M,81}=g_{M,18}\equiv g_{M}^{\prime}$ and
$g_{E,88}=g_{E,81}=g_{E,18}\equiv g_{E}^{\prime}$ . We take the
parameters $g$, $g_M$ and $g_E$ to be small and calculate SU(3)
breaking to first order in these parameters. From Eq.(\ref{19}),
Eq.(\ref{20}), the $\omega-\phi$ "ideal" mixing and the lagrangian
Eq.(22) we get the following reduced branching fractions:
\begin{eqnarray}
\nonumber&&\widetilde{Br}(J/\psi\rightarrow
E_{\pi^+}\rho^-)=\widetilde{Br}(J/\psi\rightarrow
E_{\pi^-}\rho^+)=\widetilde{Br}(J/\psi\rightarrow
E_{\pi^0}\rho^0)=|g+\frac{1}{\sqrt{3}}g_{M}^{\prime}+\frac{1}{3}g_{E}^{\prime}|^2\\
\nonumber&&\widetilde{Br}(J/\psi\rightarrow
E_{K^+}K^*-)=\widetilde{Br}(J/\psi\rightarrow
E_{K^-}K^{*-})=|g-\frac{1}{2\sqrt{3}}g_{M}^{\prime}+\frac{1}{3}g_{E}^{\prime}|^2\\
\nonumber&&\widetilde{Br}(J/\psi\rightarrow
E_{K^0}\overline{K}^{*0})=\widetilde{Br}(J/\psi\rightarrow
E_{\overline{K}^0}K^{*0})=|g-\frac{1}{2\sqrt{3}}g_{M}^{\prime}-\frac{2}{3}g_{E}^{\prime}|^2\\
\nonumber&&\widetilde{Br}(J/\psi\rightarrow
E_{\eta}\phi)=|g-\frac{2}{\sqrt{3}}g^{\prime}_M-\frac{2}{3}g^{\prime}_E|^2(\sqrt{\frac{2}{3}}\cos\varphi_P+\frac{1}{\sqrt{3}}\sin\varphi_P)^2\\
\nonumber&&\widetilde{Br}(J/\psi\rightarrow
E_{\eta}\omega)=|g+\frac{1}{\sqrt{3}}g^{\prime}_M+\frac{1}{3}g^{\prime}_E|^2(\sqrt{\frac{1}{3}}\cos\varphi_P-\sqrt{\frac{2}{3}}\sin\varphi_P)^2\\
\nonumber&&\widetilde{Br}(J/\psi\rightarrow
E_{\eta}\rho^0)=|g^{\prime}_E|^2(\sqrt{\frac{1}{3}}\cos\varphi_P-\sqrt{\frac{2}{3}}\sin\varphi_P)^2\\
\nonumber&&\widetilde{Br}(J/\psi\rightarrow
E_{\eta^{\prime}}\phi)=|g-\frac{2}{\sqrt{3}}g^{\prime}_M-\frac{2}{3}g^{\prime}_E|^2(\sqrt{\frac{1}{3}}\cos\varphi_P-\sqrt{\frac{2}{3}}\sin\varphi_P)^2\\
\nonumber&&\widetilde{Br}(J/\psi\rightarrow
E_{\eta^{\prime}}\omega)=|g+\frac{1}{\sqrt{3}}g^{\prime}_M+\frac{1}{3}g^{\prime}_E|^2(\sqrt{\frac{1}{3}}\sin\varphi_P+\sqrt{\frac{2}{3}}\cos\varphi_P)^2\\
\nonumber&&\widetilde{Br}(J/\psi\rightarrow
E_{\eta^{\prime}}\rho^{0})=|g^{\prime}_E|^2(\frac{1}{\sqrt{3}}\sin\varphi_P+\sqrt{\frac{2}{3}}\cos\varphi_P)^2\\
\nonumber&&\widetilde{Br}(J/\psi\rightarrow E_{\pi^{0}}\phi)=0\\
\label{25}&&\widetilde{Br}(J/\psi\rightarrow
E_{\pi^{0}}\omega)=|g^{\prime}_E|^2
\end{eqnarray}
From the above reduced branching fractions, the following relations
can be obtained
\begin{eqnarray}
\nonumber&&\frac{\widetilde{Br}(J/\psi\rightarrow
E_{\eta}\omega)}{\widetilde{Br}(J/\psi\rightarrow
E_{\pi^0}\rho^0)}=\frac{\widetilde{Br}(J/\psi\rightarrow
E_{\eta}\rho^0)}{\widetilde{Br}(J/\psi\rightarrow
E_{\pi^0}\omega)}=(\sqrt{\frac{1}{3}}\cos\varphi_P-\sqrt{\frac{2}{3}}\sin\varphi_P)^2\\
\label{26}&&\frac{\widetilde{Br}(J/\psi\rightarrow
E_{\eta^{\prime}}\omega)}{\widetilde{Br}(J/\psi\rightarrow
E_{\pi^{0}}\rho^{0})}=\frac{\widetilde{Br}(J/\psi\rightarrow
E_{\eta^{\prime}}\rho^0)}{\widetilde{Br}(J/\psi\rightarrow
E_{\pi^{0}}\omega)}=(\sqrt{\frac{1}{3}}\sin\varphi_P+\sqrt{\frac{2}{3}}\cos\varphi_P)^2
\end{eqnarray}
Both because $Br(J/\psi\rightarrow\omega X(1835))$ is heavily
suppressed (please see the previous section of this work) and
because there are not yet any experimental reports on it, we take
$\widetilde{Br}(J/\psi\rightarrow
E_{\eta^{\prime}}\omega)=|g+\frac{1}{\sqrt{3}}g^{\prime}_M+\frac{1}{3}g^{\prime}_E|^2(\sqrt{\frac{1}{3}}\sin\varphi_P+\sqrt{\frac{2}{3}}\cos\varphi_P)^2\approx0$,
this implies
$|g+\frac{1}{\sqrt{3}}g^{\prime}_M+\frac{1}{3}g^{\prime}_E|^2\approx0$
or
$(\sqrt{\frac{1}{3}}\sin\varphi_P+\sqrt{\frac{2}{3}}\cos\varphi_P)^2\approx0$.
If
$|g+\frac{1}{\sqrt{3}}g^{\prime}_M+\frac{1}{3}g^{\prime}_E|^2\approx0$,
we can see $\widetilde{Br}(J/\psi\rightarrow
E_{\pi^+}\rho^-)=\widetilde{Br}(J/\psi\rightarrow
E_{\pi^-}\rho^+)=\widetilde{Br}(J/\psi\rightarrow
E_{\pi^0}\rho^0)\approx0$; However if
$(\sqrt{\frac{1}{3}}\sin\varphi_P+\sqrt{\frac{2}{3}}\cos\varphi_P)^2\approx0$,
it indicates $\widetilde{Br}(J/\psi\rightarrow
E_{\eta^{\prime}}\rho^0)\approx0$, and we can not observe X(1835) in
the process $J/\psi\rightarrow\rho^0 X(1835)$, with
$X(1835)\rightarrow p\overline{p}$ or
$X(1835)\rightarrow\eta^{\prime}\pi^+\pi^-$. This conclusion is
consistent with the result $Br(J/\psi\rightarrow\rho
X(1835))<Br(J/\psi\rightarrow\omega X(1835))$ which has been
obtained in the Sec.II.B.
\subsection{$J/\psi\rightarrow E_V P$}
Similar to the above two cases, the effective lagrangian responsible
for the decay is
\begin{eqnarray}
&&\nonumber{\cal
L}_{eff}=\varepsilon_{\mu\nu\alpha\beta}F^{\mu\nu}_{\Psi}\{g^{\prime\prime}_8F^{\alpha\beta}_{E_{V_a}}P_a+g^{\prime\prime}_1F^{\alpha\beta}_{E_{\omega_1}}P_{\eta_1}+[g^{\prime\prime}_{M,88}d^{abc}F^{\alpha\beta}_{E_{V_a}}P_bM^c
+\sqrt{\frac{2}{3}}g^{\prime\prime}_{M,81}F^{\alpha\beta}_{E_{V_a}}M^aP_{\eta_1}\\
\nonumber&&+\sqrt{\frac{2}{3}}g^{\prime\prime}_{M,18}F^{\alpha\beta}_{E_{\omega_1}}M^aP_a]+[g^{\prime\prime}_{E,88}d^{abc}F^{\alpha\beta}_{E_{V_a}}P_bE^c+\sqrt{\frac{2}{3}}g^{\prime\prime}_{E,81}F^{\alpha\beta}_{E_{V_a}}E^aP_{\eta_1}
+\sqrt{\frac{2}{3}}g^{\prime\prime}_{E,18}F^{\alpha\beta}_{E_{\omega_1}}E^aP_a]\}\\
\end{eqnarray}
under the nonet symmetry, the relations
$g^{\prime\prime}_8=g^{\prime\prime}_1\equiv g^{\prime\prime}$,
$g^{\prime\prime}_{M,88}=g^{\prime\prime}_{M,81}=g^{\prime\prime}_{M,18}\equiv
g^{\prime\prime}_M$ and
$g^{\prime\prime}_{E,88}=g^{\prime\prime}_{E,81}=g^{\prime\prime}_{E,18}\equiv
g^{\prime\prime}_E$ hold. From Eq.(\ref{21}) and the above
lagrangian Eq.(25) we can obtain the following reduced branching
fractions:
\begin{eqnarray}
\nonumber&&\widetilde{Br}(J/\psi\rightarrow\pi^{+}E_{\rho^-})=\widetilde{Br}(J/\psi\rightarrow\pi^{-}E_{\rho^+})
=\widetilde{Br}(J/\psi\rightarrow\pi^{0}E_{\rho^0})=|g^{\prime\prime}+\frac{1}{\sqrt{3}}g^{\prime\prime}_M+\frac{1}{3}g^{\prime\prime}_E|^2\\
\nonumber&&\widetilde{Br}(J/\psi\rightarrow
K^{+}E_{K^{*-}})=\widetilde{Br}(J/\psi\rightarrow
K^{-}E_{K^{*+}})=|g^{\prime\prime}-\frac{1}{2\sqrt{3}}g^{\prime\prime}_M+\frac{1}{3}g^{\prime\prime}_E|^2\\
\nonumber&&\widetilde{Br}(J/\psi\rightarrow
K^{0}E_{\overline{K}^{*0}})=\widetilde{Br}(J/\psi\rightarrow
\overline{K}^{0}E_{K^{*0}})=|g^{\prime\prime}-\frac{1}{2\sqrt{3}}g^{\prime\prime}_M-\frac{2}{3}g^{\prime\prime}_E|^2\\
\nonumber&&\widetilde{Br}(J/\psi\rightarrow\eta
E_{\phi})=|g^{\prime\prime}\cos(\theta_P-\varphi_V)-(g^{\prime\prime}_M+\frac{1}{\sqrt{3}}g^{\prime\prime}_E)[\frac{1}{\sqrt{3}}\cos\theta_P\cos\varphi_V+\sqrt{\frac{2}{3}}\sin(\theta_P+\varphi_V)]|^2\\
\nonumber&&\widetilde{Br}(J/\psi\rightarrow\eta
E_{\omega})=|g^{\prime\prime}\sin(\varphi_V-\theta_P)+(g^{\prime\prime}_M+\frac{1}{\sqrt{3}}g^{\prime\prime}_E)[-\frac{1}{\sqrt{3}}\cos\theta_P\sin\varphi_V+\sqrt{\frac{2}{3}}\cos(\theta_P+\varphi_V)]|^2\\
\nonumber&&\widetilde{Br}(J/\psi\rightarrow\eta
E_{\rho^0})=|g^{\prime\prime}_E|^2(\frac{1}{\sqrt{3}}\cos\theta_P-\sqrt{\frac{2}{3}}\sin\theta_P)^2\\
\nonumber&&\widetilde{Br}(J/\psi\rightarrow\eta^{\prime}
E_{\phi})=|-g^{\prime\prime}\sin(\varphi_V-\theta_P)+(g^{\prime\prime}_M+\frac{1}{\sqrt{3}}g^{\prime\prime}_E)[-\frac{1}{\sqrt{3}}\sin\theta_P\cos\varphi_V+\sqrt{\frac{2}{3}}\cos(\theta_P+\varphi_V)]|^2\\
\nonumber&&\widetilde{Br}(J/\psi\rightarrow\eta^{\prime}
E_{\omega})=|g^{\prime\prime}\cos(\varphi_V-\theta_P)+(g^{\prime\prime}_M+\frac{1}{\sqrt{3}}g^{\prime\prime}_E)[-\frac{1}{\sqrt{3}}\sin\theta_P\sin\varphi_V+\sqrt{\frac{2}{3}}\sin(\theta_P+\varphi_V)]|^2\\
\nonumber&&\widetilde{Br}(J/\psi\rightarrow\eta^{\prime}E_{\rho^0})=|g^{\prime\prime}_E|^2(\frac{1}{\sqrt{3}}\sin\theta_P+\sqrt{\frac{2}{3}}\cos\theta_P)^2\\
\nonumber&&\widetilde{Br}(J/\psi\rightarrow\pi^{0}E_{\phi})=|g^{\prime\prime}_E|^2(\frac{1}{\sqrt{3}}\cos\varphi_V-\sqrt{\frac{2}{3}}\sin\varphi_V)^2\\
\label{28}&&\widetilde{Br}(J/\psi\rightarrow\pi^{0}E_{\omega})=|g^{\prime\prime}_E|^2(\frac{1}{\sqrt{3}}\sin\varphi_V+\sqrt{\frac{2}{3}}\cos\varphi_V)^2
\end{eqnarray}
here $\theta_P$ is the mixing angle of $\eta$ and $\eta^{\prime}$
with $\theta_P\approx-16.9^{\circ}\pm1.7^{\;\circ}$\cite{mix}, and
we can find the relation
\begin{equation}
\label{29}\frac{\widetilde{Br}(J/\psi\rightarrow\pi^0E_{\phi})}{\widetilde{Br}(J/\psi\rightarrow\pi^0E_{\omega})}=(\frac{\frac{1}{\sqrt{3}}\cos\varphi_V-\sqrt{\frac{2}{3}}\sin\varphi_V}
{\frac{1}{\sqrt{3}}\sin\varphi_V+\sqrt{\frac{2}{3}}\cos\varphi_V})^2=(\frac{1-\sqrt{2}\tan\varphi_V}{\tan\varphi_V+\sqrt{2}})^2
\end{equation}
In summary, the other exotic states in the weight diagram are
expected to be observed in future, and these relations between the
branching fractions can be served as a guide to the experimental
search for these exotic states.
\section{conclusion and discussion}
In conclusion, the processes $\Upsilon\rightarrow \gamma X(1835)$
and $J/\psi\rightarrow \omega X(1835)$ have been investigated.
Considering the large coupling of X(1835) with $p\overline{p}$ and
$\eta^{\prime}\pi^{+}\pi^{-}$,  we propose that X(1835) is a
baryonium with sizable gluon content, and mainly belongs to  a SU(3)
flavor singlet. In this scheme, we  can finely understand the
observation data  both in the process $J/\psi\rightarrow\gamma
X(1835)$, $X(1835)\rightarrow p\overline{p}$\cite{Bes2} and in
$J/\psi\rightarrow\gamma X(1835)$, $X(1835)\rightarrow
\eta^{\prime}\pi^{+}\pi^{-}$\cite{Bes1}. We estimate that in the
$\Upsilon(1S)$ radiative decay the product branching fraction
$Br(\Upsilon(1S)\rightarrow\gamma X(1835))Br(X(1835)\rightarrow
p\overline{p})<6.45\times10^{-7}$, which is compatible with the
CLEO's experimental upper limit $Br(\Upsilon(1S)\rightarrow\gamma
X(1835))Br(X(1835)\rightarrow
p\overline{p})<5\times10^{-7}$\cite{Cleo}. Noting that in these
processes the gluon component plays important role due to the
$U_A(1)$ anomaly. Thus, we find out that the drastic smallness of
$Br( \Upsilon(1S)\rightarrow\gamma X(1835))$ is caused the special
nature of $X(1835)$, and it does not contradict with the
experimental evidence of $X(1835)$ revealed in the process $J/\psi
\rightarrow\gamma X(1835)$ by BES.

In our baryonium scheme of $X(1835)$, we found out also that
$Br(J/\psi\rightarrow\omega X(1835))=(2.00\pm 0.35)\times 10^{-5} $,
$8.00\times10^{-7}<Br(J/\psi\rightarrow\omega
X(1835))Br(X(1835)\rightarrow p\overline{p})<2.80\times10^{-6},$,
and $2.40\times10^{-6}<Br(J/\psi\rightarrow\omega
X(1835))Br(X(1835)\rightarrow\eta^{\prime}\pi^{+}\pi^{-})<8.40\times10^{-6}$.
The production of X(1835) in the process $J/\psi\rightarrow\omega
X(1835)$ are heavily suppressed. We also point out that
$Br(J/\psi\rightarrow\rho X(1835))<Br(J/\psi\rightarrow\omega
X(1835))$, so it is very difficult to observe X(1835) in the process
$J/\psi\rightarrow V X(1835)$(here V is $\omega~\rm{or}~\rho$) with
$X(1835)\rightarrow p\overline{p}$ or $X(1835)\rightarrow
\eta'\pi^{+}\pi^{-}$, and the $J/\psi$ radiative decay is the most
suitable place for searching $ X(1835)$. We address that the
baryonic component dominates the decay $J/\psi\rightarrow V X(1835)$
with $V ~\rm{is}~ \omega~\rm{or}~\rho$, since the $U_A(1)$ anomaly
contributions are suppressed in these processes. The experimental
check for the above results are expected.
\par
Finally we conjecture the existence of baryonium nonet, which is
supported in Ref.\cite{yan3} and Ref.\cite{yuan}, and the nonet can
be pseudoscalar($E_{P_i}$) or vector($E_{V_i}$). The $p\overline{p}$
enhancement X(1835) is identified as $E_{\eta^{\prime}}$, and the
$p\overline{\Lambda}$ enhancement\cite{Bes5} can be $E_{K^{*+}}$ or
$E_{K^+}$. We derive the reduced branching fractions of
$J/\psi\rightarrow E_{P}P$, $J/\psi\rightarrow E_{P}V$ and
$J/\psi\rightarrow E_{V}P$ in a model independent way basing on
SU(3) symmetry with the symmetry breaking included. The relations
between the branching fractions can be served as a guide to the
experimental search for these exotic states.

\begin{center}
{\bf ACKNOWLEDGMENTS}
\end{center}
We would like to acknowledge Prof. S.Jin and Prof. X.Y. Shen for
discussion on $J/\psi\rightarrow\omega X(1835)$. This work is
partially supported by National Natural Science Foundation of China
under Grant Numbers 90403021 and 10375074, and by the PhD Program
Funds of the Education Ministry of China and KJCX2-SW-N10 of the
Chinese Academy.

\newpage
\begin{figure}[hptb]
\begin{center}
\includegraphics*[75pt,515pt][518pt,668pt]{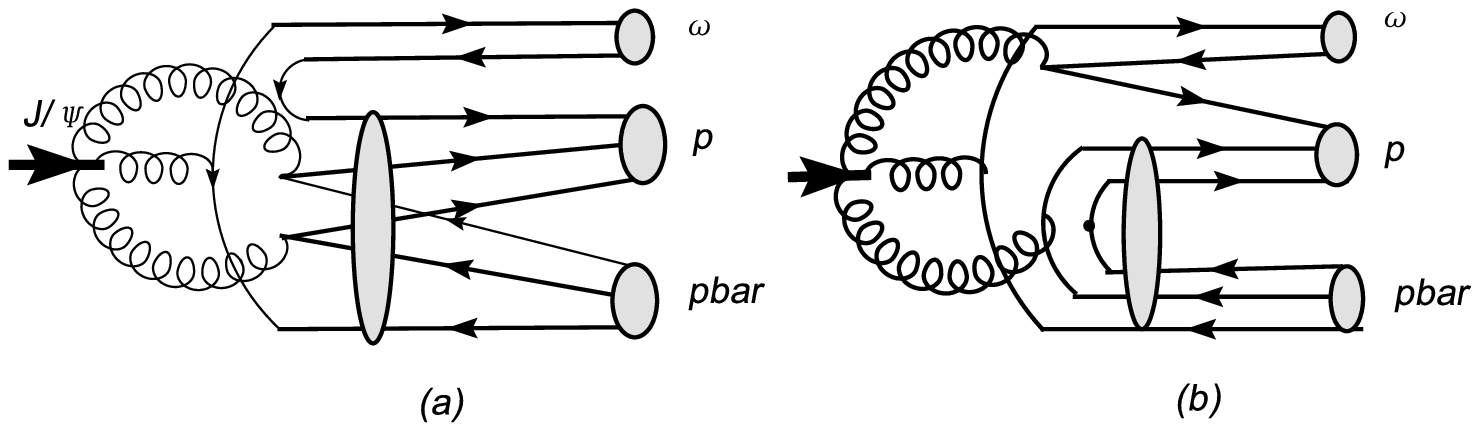}
\caption{A schematic diagram for $J/\psi\rightarrow\omega
X(1835)\rightarrow p\overline{p}$ in a mechanism with a $^3P_0$
quark pair created in two configurations.}
\end{center}
\end{figure}
\begin{figure}[hptb]
\begin{center}
\includegraphics*[100pt,560pt][493pt,713pt]{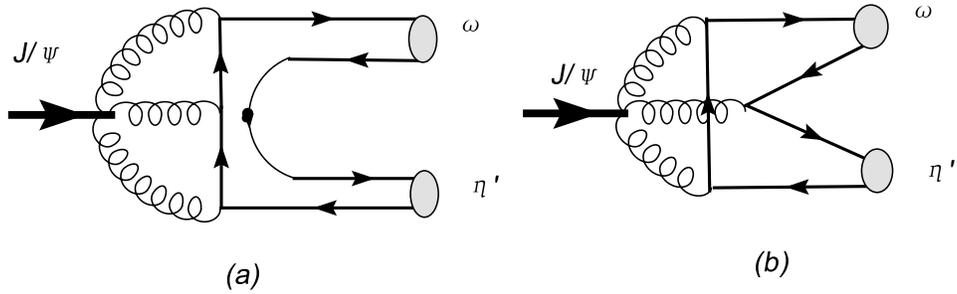}
\caption{A schematic diagram for $J/\psi\rightarrow\omega\eta'$
decay mechanism (a) a $^3P_0$ quark pair created;(b) no $^3P_0$
quark pair created.}
\end{center}
\end{figure}
\begin{figure}[hptb]
\begin{center}
\includegraphics*[100pt,590pt][290pt,760pt]{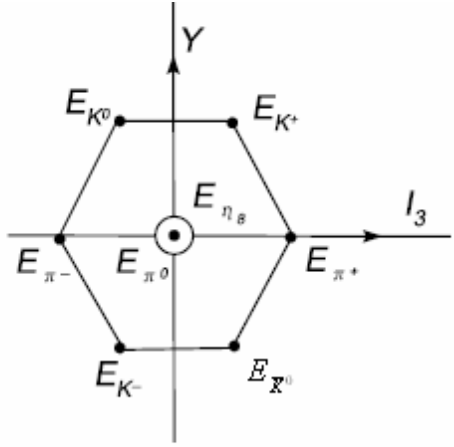}
\caption{The weight diagram for the pseudoscalar baryonium octet }.
\end{center}
\end{figure}
\begin{figure}[hptb]
\begin{center}
\includegraphics*[100pt,600pt][270pt,760pt]{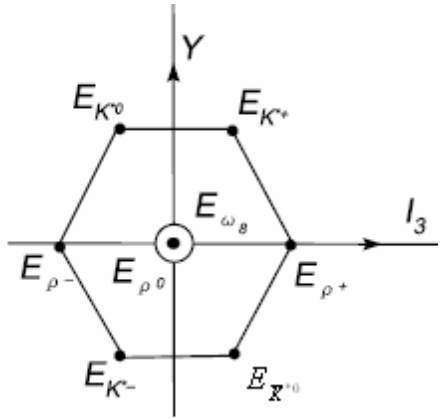}
\caption{The weight diagram for the vector baryonium octet }.
\end{center}
\end{figure}

\end{document}